\documentclass[%
reprint,
superscriptaddress,
amsmath,amssymb,
aps,
prb,
floatfix,
]{revtex4-2}

\usepackage{graphicx}% Include figure files
\usepackage{dcolumn}% Align table columns on decimal point
\usepackage{bm}% bold math
\usepackage{hyperref}% add hypertext capabilities
\usepackage[mathlines]{lineno}% Enable numbering of text and display math
\usepackage{amsmath,bm}
\usepackage[utf8]{inputenc}
\DeclareUnicodeCharacter{2212}{-}
\usepackage{times}
\usepackage{fouriernc}
\usepackage{xcolor}

\begin{document}

\preprint{APS/123-QED}

\title{Evolution of charge correlations in the hole-doped kagome superconductor CsV$_{3-x}$Ti$_x$Sb$_5$}% Force line breaks with \\

\author{Ganesh Pokharel}
\email{gpokhare@westga.edu}
\affiliation{Materials Department, University of California, Santa Barbara, California 93106, USA}
\affiliation{Perry College of Mathematics, Computing and Sciences, University of West Georgia, Carrollton, Georgia 30118, USA}

\author{Canxun Zhang}
\affiliation{Department of Physics, University of California, Santa Barbara, California 93106, USA}

\author{Evgeny Redekop}
\affiliation{Department of Physics, University of California, Santa Barbara, California 93106, USA}

\author{Brenden R. Ortiz}
\affiliation{Materials Department, University of California, Santa Barbara, California 93106, USA}

\author{Andrea N. Capa Salinas}
\affiliation{Materials Department, University of California, Santa Barbara, California 93106, USA}

\author{Sarah Schwarz}
\affiliation{Department of Physics, University of California, Santa Barbara, California 93106, USA}

\author{Steven J. Gomez Alvarado}
\affiliation{Materials Department, University of California, Santa Barbara, California 93106, USA}

\author{Suchismita Sarker}
\affiliation{Cornell High Energy Synchrotron Source, Cornell University, Ithaca, New York 14853, USA}

\author{Andrea F. Young}
\affiliation{Department of Physics, University of California, Santa Barbara, California 93106, USA}

\author{Stephen D. Wilson}
\email{stephendwilson@ucsb.edu}
\affiliation{Materials Department, University of California, Santa Barbara, California 93106, USA}

\date{\today}% It is always \today, today,

\begin{abstract}

The interplay between superconductivity and charge correlations in the kagome metal CsV$_3$Sb$_5$ can be tuned by external perturbations such as doping or pressure. Here we present a study of charge correlations and superconductivity upon hole doping via Ti substitution on the V kagome sites in CsV$_{3-x}$Ti$_x$Sb$_5$ via synchrotron x-ray diffraction and scanning SQUID measurements.  While the superconducting phase, as viewed via the vortex structure, remains conventional and unchanged across the phase diagram, the nature of charge correlations evolves as a function of hole-doping from the first superconducting dome into the second superconducting dome. For Ti doping in the first superconducting dome, competing $2\times  2 \times 2$ and $2\times  2 \times 4$ supercells form within the charge density wave state and are suppressed rapidly with carrier substitution. In the second superconducting dome, no charge correlations are detected. Comparing these results to those observed for CsV$_3$Sb$_{5-x}$Sn$_x$ suggests important differences between hole doping via chemical substitution on the V and Sb sites, particularly in the disorder potential associated with each dopant.   

\end{abstract}

\maketitle

\section{Introduction}

The kagome network, composed of corner-sharing triangles, is a structural motif that has attracted significant  interest due to its capacity for hosting inherent geometrical frustration, nontrivial topology, and strong electronic correlation effects. V-based kagome metals of the form $A$V$_3$Sb$_5$ (\textit{A} = K, Rb, Cs) or ``135'' materials present a promising materials platform to study the interplay between nontrivial band topology, superconductivity, and charge density wave (CDW) correlations \cite{PhysRevMaterials.3.094407, PhysRevLett.125.247002, Wilson2024, PhysRevMaterials.5.034801, Yin_2021, PhysRevMaterials.7.024806, PhysRevMaterials.7.014801, PhysRevMaterials.7.024806}.
Charge correlations and CDW orders within this family are hypothesized to play a key role in their anomalous properties such as time-reversal symmetry breaking \cite{Jiang2021-ay, Mielke2022-og, FENG20211384}, magnetochirality and nonreciprocal transport \cite{Wu2022, Guo2022}, pair density wave superconductivity \cite{Chen2021, PhysRevX.14.021025}, orbital magnetism \cite{yu2021evidence, Xu2022}, and beyond \cite{wilson2024v3sb5}.
%Therefore, probing the relationship between CDW orders and the superconducting state, as well as their connection to the kagome structural motif, is critical to building a comprehensive understanding of the mechanisms underlying the complex electronic phase diagram in the $A$V$_3$Sb$_5$ family.
%A comprehensive understanding of the relationship between the superconducting state and CDW orders, as well as their connection to the kagome network motif in  \textit{A}V$_3$Sb$_5$, is thus crucial to advancing the field and can be probed by using an appropriate external perturbation.
Building a comprehensive understanding of the relationship between the CDW orders and the superconducting state in $A$V$_3$Sb$_5$ is key to understanding the origin of their anomalous properties, and one approach is to analyze their evolution under chemical perturbation.

CsV$_3$Sb$_5$ exhibits the highest superconducting critical temperature in the 135 family, multi-gap features, and a double superconducting dome feature in its electronic phase diagram \cite{PhysRevLett.127.187004, Gupta2022, Neupert2022, PhysRevMaterials.6.L041801, PhysRevLett.126.247001}. 
Below $T_\mathrm{CDW} \approx 94 \mathrm{K}$, the high-temperature hexagonal lattice symmetry is lowered to an orthorhombic symmetry via a $2\times  2$ reconstruction within each kagome plane, characterized by either a Star-of-David (SoD) or tri-hexagonal (TrH) pattern \cite{PhysRevX.11.041030, PhysRevMaterials.7.024806,PhysRevB.106.L241106}. Along the out-of-plane direction, both $2\times  2 \times 2$ and $2\times  2 \times 4$ reconstructions are in close competition as observed via x-ray diffraction \cite{PhysRevMaterials.7.024806, PhysRevB.105.195136, PhysRevResearch.5.L012032}, with recent dark-field X-ray microscopy results suggesting a microscale coexistence of both phases \cite{PhysRevMaterials.8.093601}. 
Although there is a near degeneracy in the choice of interlayer stacking configuration, it can be easily influenced by small perturbations such as pressure or doping \cite{Kautzsch2023-nq, PhysRevMaterials.6.L041801, PhysRevLett.126.247001, PhysRevB.107.144502, PhysRevB.107.174107,Chen_2021}.

External perturbations such as chemical doping and mechanical pressure provide effective methods of studying the nature of the superconducting state and its relationship with CDW order. Recent studies using electrical transport measurements, magnetization, x-ray diffraction (XRD) and nuclear magnetic resonance have demonstrated the tunability of both orders in CsV$_3$Sb$_5$ under chemical doping \cite{PhysRevMaterials.6.L041801, PhysRevLett.126.247001, Kautzsch2023-nq,YANG20222176,Zheng2022_2}. 
The CDW transition is rapidly suppressed with increasing chemical doping and eventually vanishes, while superconductivity exhibits nonmonotonic behavior, forming two superconducting domes with elevated $T_\mathrm{c}$ \cite{Kautzsch2023-nq, YANG20222176, PhysRevMaterials.6.L041801}.
Hole-doping at either the V or Sb sites lowers the chemical potential in CsV$_3$Sb$_5$, modifying the relative positions of the the Van Hove singularities, strongly modifying the Sb $p_z$-pocket at $\Gamma$, and leading to a quick suppression of the CDW \cite{PhysRevMaterials.6.L041801, PhysRevB.104.205129, Kautzsch2023-nq}.  

In the case of hole-doping on the Sb sites, the charge correlations evolve into an incommensurate, quasi-1D regime beyond the phase boundary of three-dimensional (3D) CDW order, suggesting a link between the formation of two superconducting domes and a crossover in the character of the charge correlations \cite{Feng2023,Kautzsch2023-nq}. In the case of hole-doping via substitution within the kagome sites, the evolution of charge correlations is less explored in the regime where the disorder potential may be stronger to the underlying kagome-based physics \cite{PhysRevMaterials.7.064801, yang2022titanium}. As a result, exploring the relationship between the formation of the two superconducting domes and the charge correlations beyond the phase boundary of the 3D CDW order remains an important area of investigation. % This sentence should be a bit more narrow to the kagome plane

Here, we present single-crystal synchrotron X-ray diffraction measurements studying the evolution of charge correlations in Ti-doped CsV$_3$Sb$_5$. 
We examine three different compositions targeting various regions of the electronic phase diagram: (1) within the first superconducting dome (SC1), (2) along the phase boundary between SC1 and the second superconducting dome (SC2), and (3) within the SC2 dome. Similar to hole-doping at the Sb site, Ti doping drives a rapid suppression of charge ordering; however, in contrast to doping via the Sb site, incommensurate quasi-1D correlations are not observed beyond the phase boundary of the 3D CDW ordered state.  Our data suggest that stronger disorder effects realized via atomic substitution within the kagame network play an important role in the doping evolution of charge correlations in 135 compounds.  We further explore the evolution of the SC ground state upon Ti substitution via nano SQUID-on-tip measurements of the vortex lattice, revealing a conventional vortex lattice state in both SC domes.

%With light Ti doping (within the SC1 dome), 2 $\times $ 2 $\times $ 2 and 2 $\times $ 2 $\times $ 4 CDW orders initially coexist. The 2 $\times $ 2 $\times $ 4 state is more easily destabilized by continued hole doping compared to the 2 $\times $ 2 $\times $ 2 CDW state and vanishes by $x=0.05$. Along the phase boundary of the SC1 and SC2 domes, only the 2 $\times $ 2 $\times $ 2 CDW contributes to the charge correlations. Within the SC2 dome, we fail to resolve remanant charge correlations, suggesting strong disorder effects in the kagome sublattice play an important role.

\begin{figure}
\centering
\includegraphics[width=0.8\columnwidth] {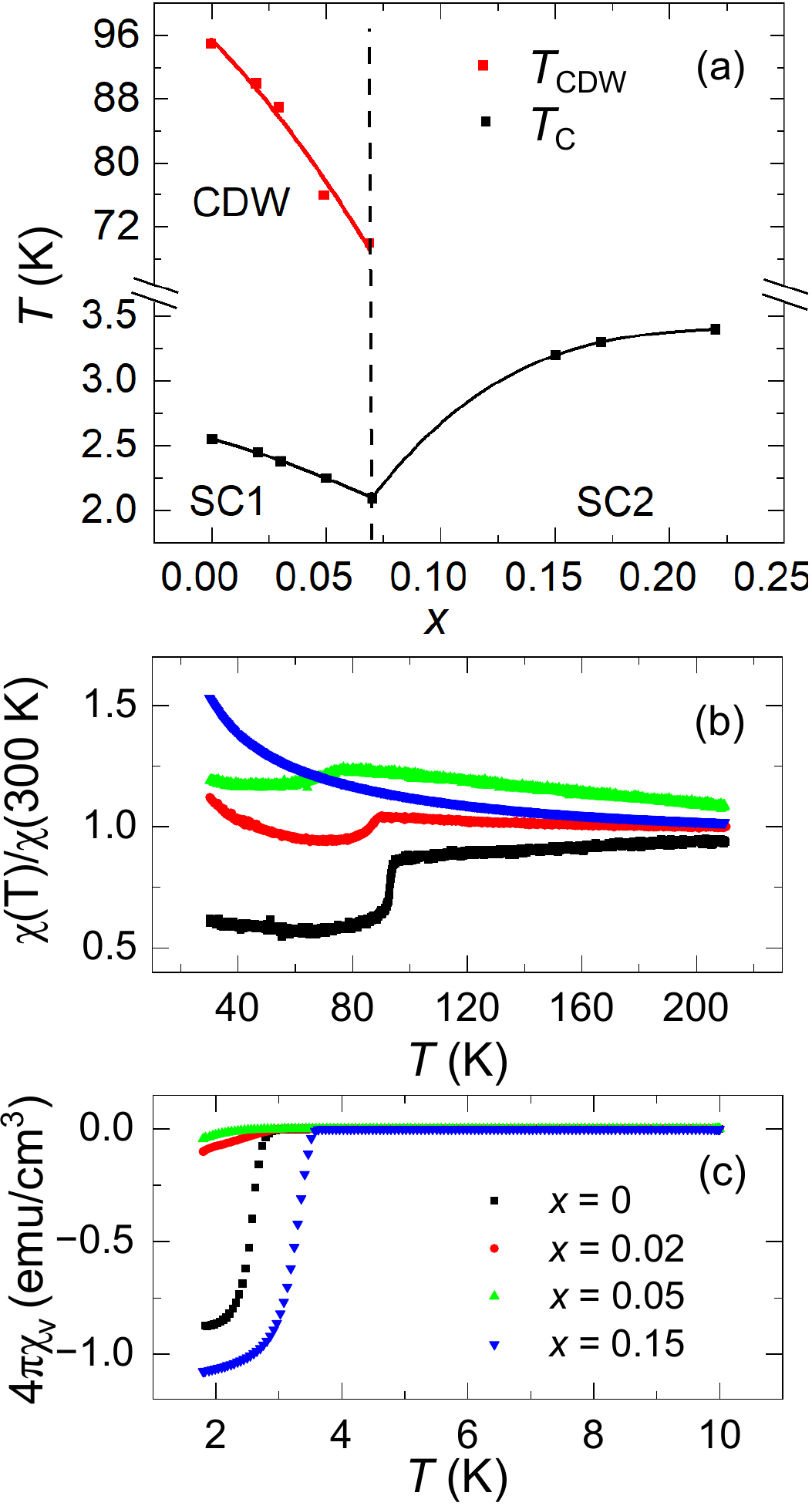}
      \caption{Evolution of electronic orders with hole doping in the kagome lattice sites. (a) Electronic phase diagram of Ti-doped CsV$_3$Sb$_5$, revealing the evolution of CDW and SC order with hole-doping in the kagome nets in single crystal samples. $T_{cdw}$ drops with the light hole-doping and completely suppresses once the second dome superconductivity appears around $x \approx 0.07$. (b, c) Comparison of the magnetic susceptibility data near the CDW and SC transitions in the parent and Ti-doped CsV$_3$Sb$_5$ crystals belonging to the SC1 and SC2 domes. }
\label{Fig_1}
\end{figure}

\begin{figure*}
\centering
\includegraphics[width=2\columnwidth] {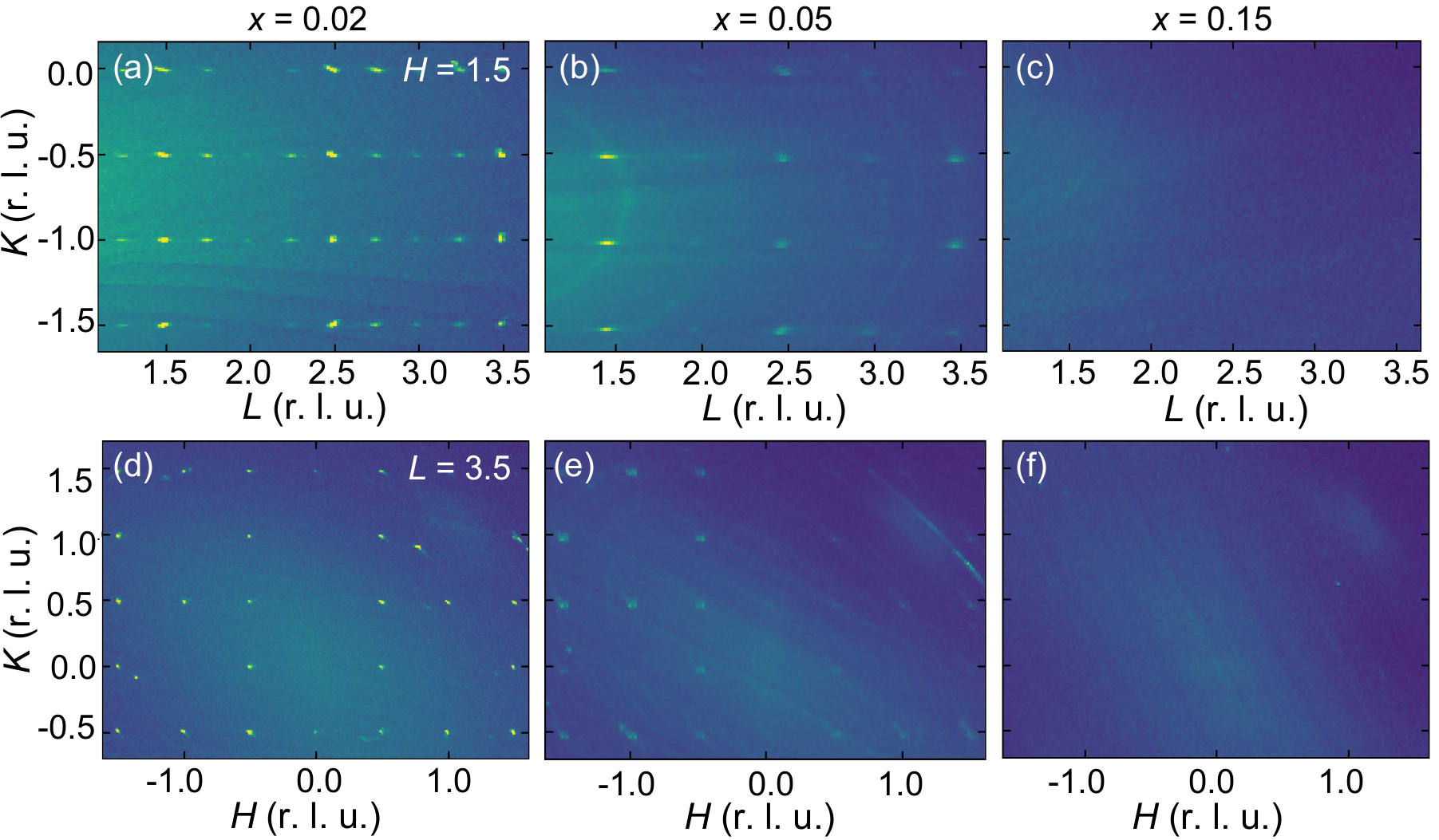}
      \caption[width=2\columnwidth]{Composition dependence of single-crystal X-ray diffraction data for CsV$_{3-x}$Ti$_x$Sb$_5$ at T=20 K. (a) $(1.5, K, L)$ scattering planes for $x=0.02$, (b) $x=0.05$, (c) $x=0.15$ crystals at $T=20$~K. (d) $(H, K, 3.5)$ scattering planes for $x=0.02$, (e) $x=0.05$, (f) $x=0.15$ crystals at $T=20$~K.}
\label{Fig_2}
\end{figure*}

\section{Experimental Details}
\subsection{Crystal synthesis}
Single crystals of CsV$_{3-x}$Ti$_x$Sb$_5$ were grown via a flux-based growth technique. V (powder, Sigma 99.9\%) was first cleaned using a mixture of hydrochloric acid and isopropyl alcohol to remove residual oxides. Cs (liquid, Alfa 99.98\%), V, Ti (powder, Alfa 99.9\%), and Sb (shot, Alfa 99.999\%) were loaded inside a pre-seasoned tungsten carbide milling vial with various molar ratios and then sealed in an Ar atmosphere. To grow the Ti-doped samples of CsV$_{3-x}$Ti$_x$Sb$_5$ ($x=0$, 0.02, 0.03, 0.05, 0.07, 0.15, 0.17, 0.22), elemental stoichiometries were selected, respectively, as Cs$_{20}$V$_{15-x}$Ti$_x$Sb$_{120}$ ($x=0$, 0.33, 0.5, 1, 2, 2.5, 3, 4) and then milled for one hour. The milled powder was poured into a cylindrical alumina crucible and then sealed inside a stainless steel tube under Ar atmosphere. The samples were heated at a rate of 200$^\circ$~C/h to 1000$^\circ$~C and dwelled for 10~h. The tubes were then cooled to 900$^\circ$~C at 25$^\circ$~C/h and to 500$^\circ$~C at 1$^\circ$~C/h. Once cooled, plate-like, shiny single crystals were manually separated from the flux and cleaned with ethanol.  

\subsection{Bulk characterization}

The doping concentration of Ti in CsV$_{3-x}$Ti$_x$Sb$_5$ single crystal samples was measured using a table top scanning tunneling microscope (Hitachi TM4000Plus) using energy dispersive spectroscopy.
Magnetic susceptibility measurements were performed as a function of temperature and magnetic field using a Quantum Design Magnetic Property Measurement System (MPMS3). To remove any surface contamination before the measurements, the surfaces of the single crystals were cleaved gently using tape. Crystals of mass $\approx$ 2~mg were attached to quartz paddles using GE varnish with the flat surface pointing towards the wall of the paddle and then loaded in the MPMS3. Magnetization data were collected in the temperature range of $T=1.8$~K to $T=300$~K with a magnetic field of $\mu_0 H = 10$~Oe and 1~T applied parallel to the crystal surface (the $ab$-plane). 

\subsection{Scattering measurements}
 
Synchrotron X-ray diffraction experiments on CsV$_{3-x}$Ti$_x$Sb$_5$ ($x=0.02$, 0.05, 0.15) crystals were performed on the ID4B (QM2) beamline at the Cornell High Energy Synchrotron Source (CHESS). An incident X-ray of wavelength  $\lambda = 0.676~\AA$ ($E = 26$~keV) was selected to avoid the absorption edges of constituent elements using a double-bounce diamond monochromator.  
The temperature of the sample was controlled by a stream of both nitrogen and cold He gas flowing across the crystal. % updated 6/9/25 SGA
The X-ray scattering data were collected in transmission geometry using a Pilatus 6M area detector array. Samples were rotated with three tilted 360$^\circ$ rotations, sliced into 0.1$^\circ$ frames, yielding 3600 data frames in each rotation. The data were visualized and analyzed using the NeXpy software package \cite{Konnecke:po5029}.

\subsection{nSOT measurements}
For nano SQUID-on-tip (nSOT) measurements, a sharp tip with an apex diameter of $\approx~150$~nm to 300~nm was fabricated from a quartz micropipette and Ti/Au films were deposited on its sides via electron beam evaporation. An additional Ti/Au strip deposited at around 0.5~mm results in a shunt resistor of $\sim$ 1~$\Omega$ to 5~$\Omega$. Indium was then evaporated in a home-built thermal evaporator held at $T < 20$~K from three angles to cover two contacts at 110$^\circ$ to the apex, with a head-on deposition performed last. This allows for a highly uniform, low-grain-size film to form near the tip apex, with two opposite weak links forming the two Josephson junctions of the superconducting quantum interference device (SQUID). Local magnetometry measurements were performed in a home-built wet fridge with a base temperature of $T\approx1.6$~K, with the SQUID placed in a quasi-voltage bias configuration and the average out-of-plane magnetic field at the tip apex read out through a series SQUID array amplifier (SSAA) \cite{huber_dc_2001}. Measurements were performed either in the direct current (dc) mode to acquire the spatially dependent $B_\perp$, or in the spatial gradient mode where a tuning fork was attached to the tip and the alternating current (ac) magnetic response at the tuning fork resonance frequency ($\approx$32~kHz) was acquired, \textit{i.e.}, $\delta B_\perp = \mathbf{a} \cdot \nabla_\mathbf{a}B_\perp$ where $\mathbf{a}$ is the tuning fork oscillation. The spatial gradient mode is preferred when the magnetic signal is relatively weak to suppress the $1/f$ noise as well as long-term drift.

% During measurement, the SQUID is placed in a quasi-voltage bias configuration and the average magnetic field at the tip apex is read out through a series SQUID array amplifier (SSAA). Measurements were performed either in the direct current mode to acquire the spatially dependent $B_\mathrm{DC}$ or in the spatial gradient mode where a tuning fork was attached to the tip and the alternating current magnetic response at the tuning fork resonance frequency measured, \textit{i.e.}, $\delta B$ = a·$\nabla {a} B$ where a is the tuning fork oscillation.

\section{Results}

To understand the evolution of charge correlations upon hole-doping in the kagome lattice sites, three different Ti concentrations were chosen across the electronic phase diagram of CsV$_{3-x}$Ti$_x$Sb$_5$. The phase diagram determined by magnetization measurements is shown in Fig. \ref{Fig_1}(a), consistent with earlier reports \cite{PhysRevMaterials.7.064801, yang2022titanium}. The first two concentrations, $x=0.02$ and $x=0.05$, fall within the SC1 dome of the phase diagram and exhibit both superconductivity (SC) and observable CDW transitions, as illustrated in Fig. \ref{Fig_1}(b, c). The temperature-dependent magnetic susceptibility, $\chi(T)$, in Fig. \ref{Fig_1}(b, c) reveals that both SC and CDW transition temperatures are suppressed with hole doping from $x=0$ to $x=0.05$. A distinct feature observed in the SC1 dome is the decrease in the superconducting volume fraction with the introduction of disorder in the kagome network.

The third concentration, $x= 0.15$ falls within the SC2 dome and retains the SC transition, recovering a full volume fraction [Fig. \ref{Fig_1}(c)]. The thermodynamic signature of CDW order in susceptibility, however, is absent [Fig. \ref{Fig_1}(b)]. The appearance of the double-dome feature in the phase diagram and the emergence of this second SC dome concomitant with the suppression of CDW order is consistent with hole doping on the Sb sites in this system \cite{PhysRevMaterials.6.L041801, Kautzsch2023-nq,10.3389/femat.2023.1257490}, though the suppression of the superconducting volume fraction and the lowering of $T_c$ with light Ti doping is a distinct behavior not observed with doping on Sb sites \cite{PhysRevMaterials.6.L041801, Kang2023}. 

X-ray diffraction data as a function of Ti-doping in CsV$_{3-x}$Ti$_x$Sb$_5$ compounds are shown in Fig. \ref{Fig_2}(a-f). Figures \ref{Fig_2}(a) and \ref{Fig_2}(d) present the $(1.5, K, L)$ and $(H, K, 3.5)$ scattering planes at $T=20$~K for the lowest Ti doping of $x=0.02$. Bragg reflections centered at $(H, K)$=$\left(\frac{1}{2}, \frac{1}{2}\right)$-type positions indicate that the parent ($2\times 2$) CDW order along the in-plane direction remains in the $x=0.02$ compound. Similarly, the parent out-of-plane correlations are also evidenced by $\frac{1}{2}$-type and $\frac{1}{4}$-type reflections along the $L$-direction in the $x=0.02$ sample. 
%Fig. \ref{Fig_2}(a,d) provides a narrower field of view to demonstrate that CDW peak positions and intensities are symmetric with respect to $\pm L$. % Why is this sentence here? Doesn't look like that's provided. Or maybe he's referencing the linecut?
Weak reflections also appear at integer $L$ positions, similar to the case for the parent CsV$_3$Sb$_5$ material \cite{kautzsch2023structural, li2022discovery}.
Unlike CsV$_3$Sb$_{5-x}$Sn$_x$, where light Sn doping with $x = 0.025$ alters the $2\times  2 \times 4$ supercell correlations \cite{Kautzsch2023-nq}, a similar hole doping level introduced via substitution on the kagome lattice sites fully preserves mixed CDW state of the parent compound. .

\begin{figure}
\centering
\includegraphics[width=0.8\columnwidth] {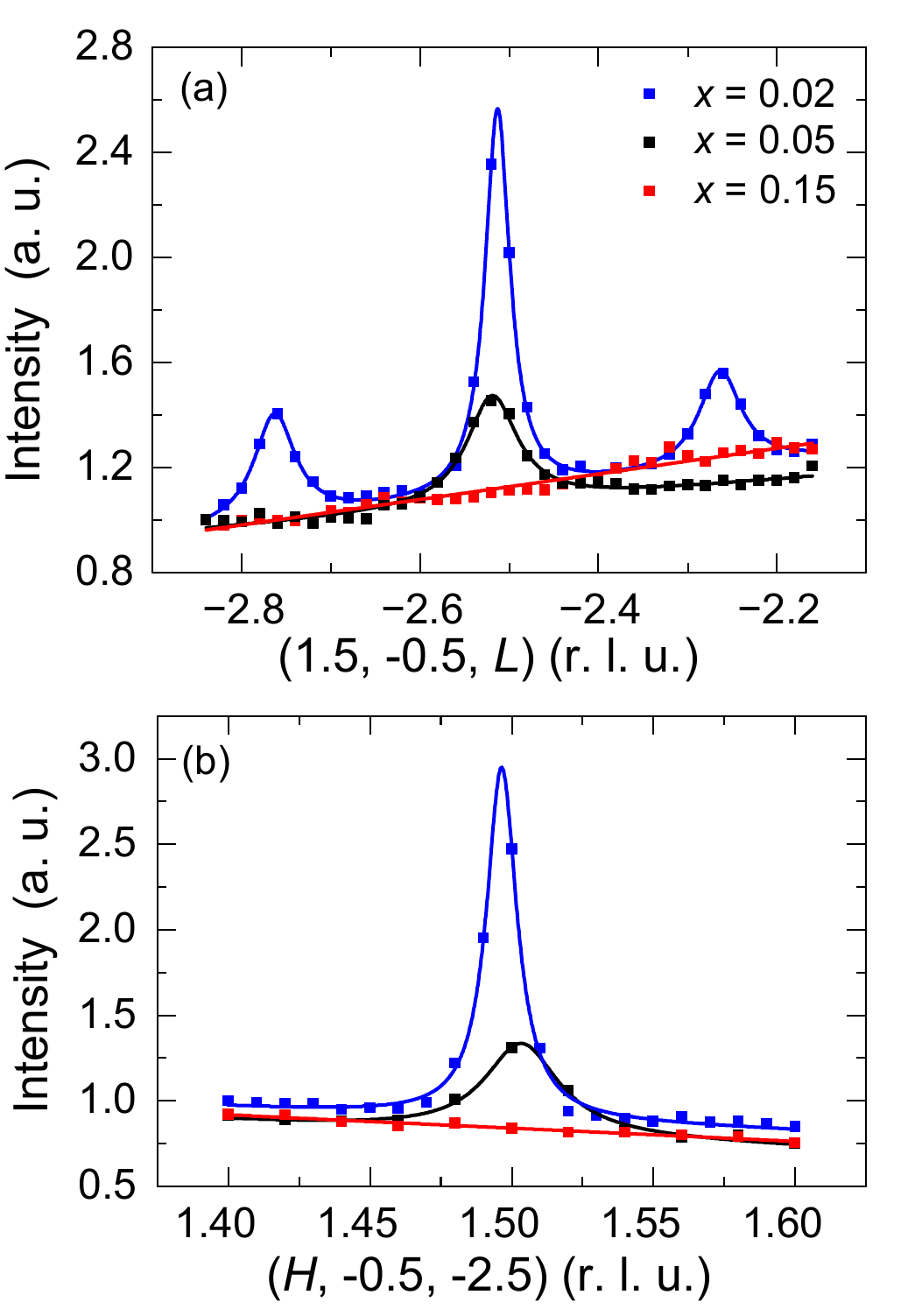}
      \caption{(a) One-dimensional line-cuts of single-crystal X-ray scattering data at $T=20$~K for CsV$_{3-x}$Ti$_x$Sb$_5$ along $(1.5, -0.5, L)$ and (b) along $(H, -0.5, -2.5)$ for $x$ = 0.02, 0.05, and 0.15 samples. Solid lines depict fits to the peak intensity using pseudo-Voigt line shapes, keeping the Gaussian component fixed to the instrumental resolution.}
\label{Fig_3}
\end{figure}

\begin{figure}
\centering
\includegraphics[width=1\columnwidth] {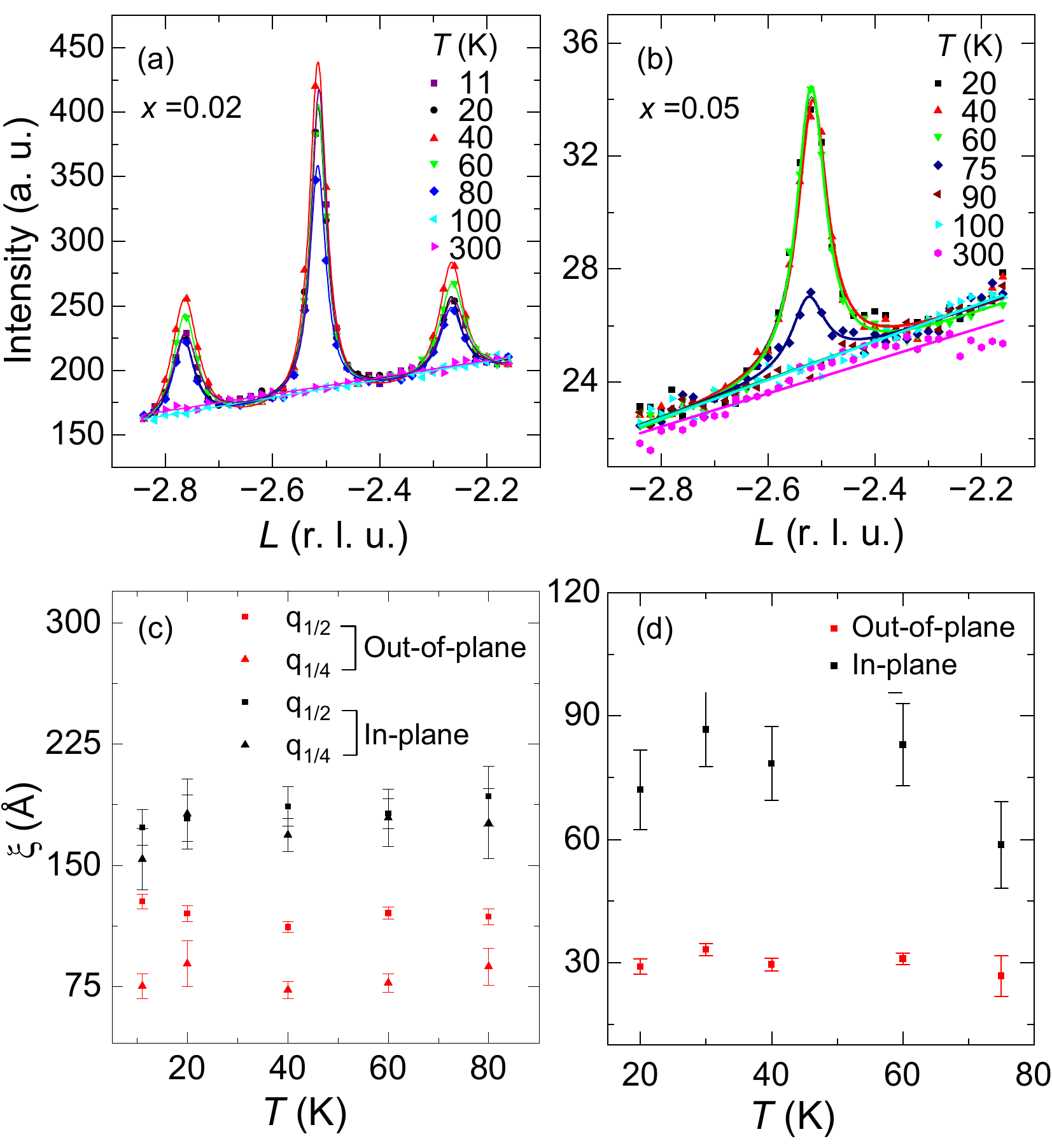}
      \caption{(a) Temperature dependence of X-ray scattering intensity at $(1.5, -0.5, -L)$ positions for $x=0.02$ and (b) $x=0.05$. (c) Temperature evolution of the in-plane and out-of-plane correlation lengths inferred from both $\frac{1}{2}$- and $\frac{1}{4}$-type reflections for $x=0.02$ and (d) $x=0.05$.}
\label{Fig_4}
\end{figure}

\begin{figure*}
\centering
\includegraphics[width=2\columnwidth] {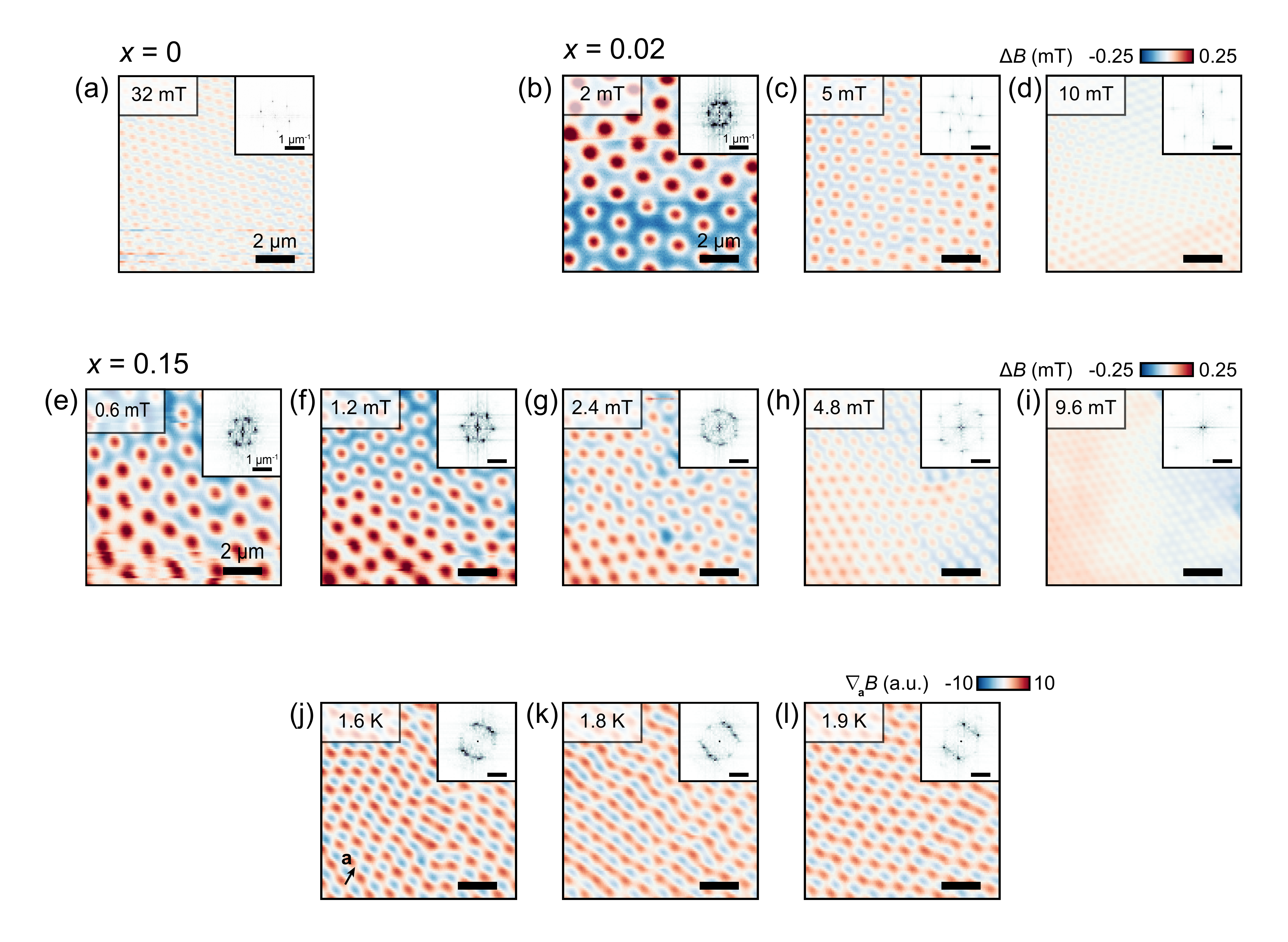}
      \caption[width=2\columnwidth]{Magnetic images of the vortex lattice in CsV$_{3−x}$Ti$_x$Sb$_5$. Panels (a--d) show the spatial map of the out-of-plane magnetic field along the $ab$ surface of crystals lying in the SC1 dome.
      % field distribution along the $ab$ surface of crystals lying in the SC1 dome. 
      (a) Magnetic image of the parent $x=0$ compound at an applied magnetic field of $\mu_0H=32$~mT and temperature of $T=1.6$~K. 
      (b-d) Magnetic images of the $x=0.02$ compound displaying vortex lattices at magnetic fields of $\mu_0H = 2$~mT, 5~mT, and 10~mT and temperature of $T=1.6$~K. 
      Panels (e--l) display the magnetic images for $x=0.15$, with the crystal lying within the SC2 dome of CsV$_{3-x}$Ti$_x$Sb$_5$ phase diagram. 
      (e--i) Images of vortex lattices with an increasing field from $\mu_0H = 0.6$~mT to 9.6~mT and temperature of $T=1.6$~K, and 
      (j--l) images of vortex lattices (spatial contrast mode) with increasing temperatures from $T=1.6$~K to $T=1.9$~K at $\mu_0H = 2.4$~mT.}
\label{Fig_5}
\end{figure*}

For samples with Ti doping $x=0.05$, the $(1.5, K, L)$ and $(H, K, 3.5)$- scattering planes at $T=20$~K are presented in Fig. \ref{Fig_2}(b) and \ref{Fig_2}(e), respectively. The scattering intensities centered at $\frac{1}{2}$-type positions in the $(H, K)$-plane reveal the persistence of the ($2\times 2$) in-plane CDW order in the $x = 0.05$ sample. However, the absence of the $\frac{1}{4}$-type of reflections along $L$ in this sample indicates the suppression of the $2\times  2 \times 4$ CDW order near the phase boundary of the SC1 and SC2 domes. Additionally, the relatively weak $\frac{1}{2}$-type SL reflections reflect the fragility of the $2\times  2 \times 2$ correlations upon hole doping in CsV$_{3-x}$Ti$_x$Sb$_5$. % Again, check electron count
The disappearance of $L=\frac{1}{4}$-type peaks near this phase boundary suggests a suppression in the mixed character of the CDW correlations (modulated SoD and TrH orders) and a crossover into a staggered tri-hexagonal-like regime with phase-shifted planes of a single distortion type, similar to (K,Rb)$_3$Sb$_5$ \cite{PhysRevMaterials.7.024806}. 
% Near the phase boundary, Ti-doped CsV$_3$Sb$_5$ shows consistent behavior to undoped members K$/$RbV$_3$Sb$_5$ by suppressing 2 $\times $ 2 $\times $ 4 correlations and resulting a single CDW phase whose ordering vector matches the other undoped members of the AV$_3$Sb$_5$ family. % This is the same sentence

Upon doping further, for Ti concentration $x=0.15$, which is beyond the nominal $2 \times 2$ CDW phase boundary, X-ray scattering data for the $(1.5, K, L)$ and $(H, K, 3.5)$ planes at $T=15$~K are showm in Fig. \ref{Fig_2}(c,f). In these data there is no evidence for charge correlations or superlattice reflections, indicating the suppression of all CDW correlations native to the SC1 dome. 

The evolution of CDW order is further illustrated via one-dimensional (1D) line-cuts at $T=20$~K shown in Fig. \ref{Fig_3}(a, b). Along the out-of-plane direction Fig. \ref{Fig_3}(a), the $\frac{1}{2}$-type SL peaks are much stronger than the $\frac{1}{4}$-type SL peaks in the $x=0.02$ sample, suggesting that the $2\times  2 \times 2$ CDW order is more stable than the $2\times  2 \times 4$ state with light Ti doping. As the doping level increases, the $\frac{1}{4}$-type peaks vanish and the $\frac{1}{2}$-type peaks also become relatively weaker in the $x=0.05$ sample. Ultimately all correlations disappear in the $x=0.15$ sample. Fig. \ref{Fig_3}(b) shows the same trend in charge correlations via in-plane cuts through the CDW superlattice positions. 

The temperature dependence of the SL peak intensities at $\frac{1}{2}$-type and $\frac{1}{4}$-type positions, and the corresponding CDW correlation lengths for $x=0.02$ and $x=0.05$ samples are shown in Fig. \ref{Fig_4}. Fig. \ref{Fig_4}(a) shows 1D line-cuts 
about the (1.5, -0.5, -2.75), (1.5, -0.5, -2.25) ($\frac{1}{4}$-type along $L$) and (1.5, -0.5, -2.5) ($\frac{1}{2}$-type along $L$) SL reflections at various temperatures for the $x=0.02$ sample. The peak intensities at $T=80$~K and below are saturated, suggestive of a first-order onset of charge correlations for the $2\times  2 \times 2$ and $2\times  2 \times 4$ CDW states. Fig. \ref{Fig_4}(b) shows a similar behavior for $2\times  2 \times 2$ type correlations for the $x=0.05$ sample.  The onset temperatures are both roughly consistent with those determined with the anomalies in the magnetization data presented in Fig. 1 (a) as well as those reported in previous studies \cite{PhysRevB.107.144502, PhysRevMaterials.7.064801, yang2022titanium}.

The evolution of the CDW correlation lengths with temperature for $x=0.02$ and $0.05$ was parameterized by resolution-convolved Lorentzian fits to the in-plane and out-of-plane SL peak profiles. Fits to the line-cuts are represented by the solid line curves in Fig. \ref{Fig_4}(a,b). The instrumental resolution combined with the crystal's intrinsic crystallinity was first determined by fitting to a primary Bragg reflection. Then, the correlation length was determined by fitting peaks to pseudo-Voigt line shapes after fixing the width of the Gaussian component to the resolution estimated from the nearby Bragg reflection.  The full width at half maximum $\omega$ (in units of~\AA$^{-1}$) of the convolved Lorentzian lineshape was used to calculate the correlation length $\xi=\frac{2}{\omega}$. 

The correlation lengths obtained by fitting the SL peaks are plotted in Fig. \ref{Fig_4}(c, d). The in-plane correlation length, $\xi_{ab}$, is nearly twice as long as the out-of-plane correlation length, $\xi_c$, which is typical for the $A$V$_3$Sb$_5$ family \cite{Kautzsch2023-nq, PhysRevX.11.041030}. The $\xi_{ab}$ associated with both $\frac{1}{2}$-type and $\frac{1}{4}$-type CDW peaks in the $x= 0.02$ sample is reduced by nearly half compared to the parent compound \cite{Li2022-kr, PhysRevX.11.041030}, shortening from resolution-limited in the undoped material to $\xi_{ab} \approx 200 \pm 20~\AA$. The out-of-plane correlation length is shorter for the $2\times  2 \times 4$ state compared to the $2\times  2 \times 2$ state, consistent with the notion that the $2\times  2 \times 4$ state is more easily suppressed upon hole doping. This is further confirmed via cuts through SL reflections for the $x=0.05$ sample where the $2\times  2 \times 4$ is absent and both the in-plane and out-of-plane correlations lengths are shortened for the remnant $2\times  2 \times 2$ state. As a function of temperature, the correlation lengths for both compounds are relatively unchanged upon cooling deeper into their respective CDW states, and charge correlations become more anisotropic (quasi-2D) with continued hole-doping.  

% Lots of ways to denote half and quarter integer here (L/2, half-integer, 1/2-type, etc.)... they're talking exclusively along L so should pick one format.

\subsection{Vortex lattice measurements}

To explore the evolution of the superconducting state across the hole-doped phase diagram, magnetic imaging of the vortex lattice in CsV$_{3-x}$Ti$_x$Sb$_5$ was performed using a nanoscale superconducting quantum interference device on the tip (nano SQUID-on-tip, nSOT) sensor. The magnetic images of the vortex lattices in CsV$_{3-x}$Ti$_x$Sb$_5$ ($x=0$, 0.02, 0.15) compounds are shown in Fig. \ref{Fig_5}. For the parent compound ($x=0$), the out-of-plane magnetic field map at $T=1.6$~K shows a conventional Abrikosov triangular vortex lattice with $h/2e$ vortices. No sign of fractional vortices were observed corresponding to higher than $2e$ pairing or composite pairing states.

For both $x = 0.02$ and $x = 0.15$ samples, the Abrikosov vortices continue to show conventional triangular lattices, whose periodicity is roughly inversely proportional to the square root of the applied out-of-plane magnetic field. As the field is lowered (\textit{i.e.}, the lattice becomes sparser), dislocations are often observed that interrupt the regular triangular pattern, as is also evidenced in the blurred peaks in the Fourier-transformed images. This likely arises from the pinning of individual vortices by local disorder. At very low fields the lattice becomes even more unstable, resulting in switching behavior [Fig. \ref{Fig_5}(b,e,f)]. Increasing the temperature from $T=1.6$~K to $T=1.9$~K can partially suppress spatial dislocations by enhancing the mobility of vortices, but the spatial contrast is no longer discernible above $T=2$~K despite the enhanced $T_\mathrm{c}$s of the doped samples.
% the lattice disappears altogether above $T = 2$~K despite the enhanced $T_c$s of the doped samples. 

\section{Discussion} 

The differences in the evolution of charge correlations between CsV$_{3-x}$Ti$_x$Sb$_5$ relative to CsV$_3$Sb$_{5-x}$Sn$_{x}$ \cite{Kautzsch2023-nq} suggest that the nature of dopant disorder has an impact on the stability of competing charge correlations.   In both compounds, light hole-doping suppresses the $2\times  2 \times 4$ CDW state faster than the $2\times  2 \times 2$ state in a qualitatively similar fashion; however the $2\times  2 \times 4$ state is slightly more robust to hole-substitution in CsV$_{3-x}$Ti$_x$Sb$_5$.  This is potentially due to the key role of Sb $p_z$ orbitals in stabilizing the 3D CDW state \cite{PhysRevB.105.235145, PhysRevB.108.075102, alkorta2025symmetrybrokenchargeorderedgroundstate} and the stronger impact on longer-range correlations upon alloying directly into the Sb sites with Sn.   The delay in the complete suppression of the $2\times  2 \times 4$ state may also explain the failure to form a peak under finite doping in SC1 for CsV$_{3-x}$Ti$_x$Sb$_5$ relative to the two-peaks at finite doping in the double-dome structure of CsV$_3$Sb$_{5-x}$Sn$_{x}$. 

A major difference between Ti-doping and Sb-doping is the absence of remnant charge correlations in the second superconducting dome, SC2, for the former case.  The quasi-1D short-range correlations that persist in SC2 \cite{Kautzsch2023-nq} are extremely weak and an increased disorder potential may push them below the detection limit of the current X-ray measurements.  If true, then the correlations in this regime originate from the kagome sublattice and direct alloying of V with Ti broadens/weakens the competing incommensurate charge correlations substantially. Alternatively, the competing charge correlations observed in  CsV$_3$Sb$_{5-x}$Sn$_{x}$ may be completely suppressed due to changes in the overall electronic band structure.  In this scenario, band structure details such as the positions of VHSs relative to $E_F$ differ between the two cases of Ti- and Sn-doping. Future, high-resolution ARPES measurements comparing the band structure differences under comparable hole-doping values will be necessary to constrain this possibility.  The recent observation of a competing CDW state beyond the suppression of $2\times  2 \times 2$ correlations in a hydrostatic pressure study \cite{PhysRevLett.133.236503} suggests that disorder effects obscuring the remnant correlations are the more likely possibility.          

nSOT measurements within the SC state reveal a conventional Abrikosov vortex lattice for all doping concentrations.  Notably, the SC state shows only conventional $h/2e$ vortices, which precludes the possibility of composite $6e$ or $4e$ pairing in the ground state.  Recent Little-Parks measurements have suggested that such a composite state may form in the fluctuation regime of CsV$_3$Sb$_5$ \cite{PhysRevX.14.021025}.  Our current measurements cannot test this fluctuating regime as the magnetic signal for the vortex lattice above 2~K falls below the sensitivity of our sensor, likely due to either reduced spatial contrast or large temperature fluctuations; however, the data can constrain theories requiring fractional vortices in the ground state. Further nSOT measurements of vortex states in ring devices may help constrain models further.

\section{Conclusions} 

In conclusion, our results provide a comparison between the landscape of charge correlations in the hole-doped kagome system CsV$_{3-x}$Ti$_x$Sb$_5$ to previous results reported in CsV$_3$Sb$_{5-x}$Sn$_{x}$ \cite{Kautzsch2023-nq}. Both $2\times  2 \times 2$ and $2\times  2 \times 4$ CDW supercell correlations exist below $T_\mathrm{CDW}$ in lightly hole-doped CsV$_{3-x}$Ti$_x$Sb$_5$ $(x=0.02)$ and the metastable $2\times  2 \times 4$ CDW state vanishes near the phase boundary between the SC1 and SC2 domes ($x=0.05$). Continued hole doping into the SC2 dome ($x=0.15$) suppresses all charge correlations with no remnant, short-range order resolved, differing from the case of CsV$_3$Sb$_{5-x}$Sn$_{x}$. Our findings provide important experimental insights into the evolution of charge correlations driven by hole doping into CsV$_3$Sb$_{5}$ and stress the importance of the chemical nature of the dopant and its corresponding disorder potential.

\section{Acknowledgments}

% Make sure we use the new MRL and QF things
This work was supported by the U.S. Department of Energy (DOE), Office of Basic Energy Sciences, Division of Materials Sciences and Engineering under grant no. DE-SC0020305.  The research made use of the facilities established by the National Science Foundation (NSF) through Enabling Quantum Leap: Convergent Accelerated Discovery Foundries for Quantum Materials Science, Engineering and Information (Q-AMASE-i): Quantum Foundry at UC Santa Barbara (DMR-1906325) and the NSF Materials Research Science and Engineering Center at UC Santa Barbara (DMR-2308708). 
 Research conducted at the Center for High-Energy X-ray Science (CHEXS) is supported by the National Science Foundation (BIO, ENG and MPS Directorates) under award DMR-2342336.

%\bibliographystyle{}
%\printbibliography
\bibliographystyle{apsrev4-2}
\bibliography{bib.bib}

\end{document}